\documentclass[pop,fleqn]{w-art}
\usepackage{times,cite,w-thm}
\theoremstyle{plain}

\theoremstyle{definition}

\usepackage[]{graphicx}
\usepackage{amsmath}
\usepackage{amsfonts}
\usepackage{amssymb}
\usepackage{calc}
\usepackage{color}
\usepackage{sidecap}
\usepackage{caption2}

\newcommand{\Nugual}[1]{$\mathcal{N}= #1 $}

\begin{document}
\pagespan{1}{}



\title[Unquenched flavors in the Klebanov-Strassler theory]{Unquenched flavors in the Klebanov-Strassler theory}


\author[S. Cremonesi]{Stefano Cremonesi \inst{1,}%
  \footnote{E-mail:~\textsf{cremones@sissa.it},
            Phone: +0039\,040\,3787\,350,
            Fax: +0039\,040\,3787\,528}}
\address[\inst{1}]{SISSA/ISAS and INFN, sez. di Trieste\\
via Beirut 2-4, 34014 Trieste (ITALY)}
\begin{abstract}
We present an analytic solution of type IIB supergravity plus D7-branes, describing the addition of any  number of flavors to the Klebanov-Strassler background. The dual field theory and its self-similar RG flow, described by a cascade of Seiberg dualities also in the presence of flavors, are discussed. The solution indicates that the dual gauge theory enjoys a duality wall in the UV. 
We stress the correspondence between Seiberg duality on the field theory side, and large gauge transformations on the RR and NSNS potentials on the gravity side. 
This contribution is mostly based on \cite{Benini:2007gx}.

\end{abstract}
\maketitle                   





\section{Introduction}

Since the seminal work of 't Hooft \cite{'tHooft:1973jz} in the Seventies, large $N$ expansions have been singled out as valuable analytic tools for understanding strongly coupled gauge theories like QCD. AdS/CFT duality, conjectured by Maldacena \cite{Maldacena:1997re}, has provided for the first time a resummation of all the planar diagrams of a strongly coupled 4-dimensional gauge theory by means of an explicit critical string dual. After the duality proposal, much effort has been made in order to extend it to less supersymmetric and nonconformal, possibly confining gauge theories. In this proceeding, we will start from a very well known example of that, the Klebanov-Strassler (KS) solution \cite{Klebanov:2000hb}, which, together with a UV completion, describes the low energy dynamics of  \Nugual{1} $SU(M)$ pure SuperYangMills (SYM) at large $M$.

\Nugual{1} pure SYM is in a loose sense halfway between nonsupersymmetric pure YM theory and QCD, having fermions in addition to gauge bosons; differently from QCD, these fermions are Majorana and transform in the adjoint representation of the gauge group. As a consequence, there is no flavor structure, and the theory is linearly confining (area law for Wilson loops), as occurs in nonsupersymmetric pure YM.

With the aim of getting closer to QCD and learning about hadronic physics, adding to the AdS/CFT correspondence flavor fields (transforming in the fundamental + antifundamental representation of the gauge group)  is of obvious importance. Karch and Katz \cite{Karch:2002sh} have studied the addition of noncompact probe branes (flavor branes) to Maldacena's setup: open strings stretching between the color and flavor branes represent the new flavor fields. In the decoupling limit of color branes from gravity, the original flavor-flavor open strings decouple too, whereas the original flavor-color open strings (quarks) become dual to new flavor-flavor open strings, which represent the excitation of flavor mesonic operators in the field theory.
One therefore needs to consider the addition of an open string sector hosted by noncompact flavor branes in the background generated by the color branes after decoupling them from gravity.

In the approach of \cite{Karch:2002sh}, flavor branes are treated as probes in the geometry sourced by the color branes, and meson spectra can be computed by studying fluctuations of the flavor brane embedding \cite{Kruczenski:2003be}. Neglecting the backreaction of the flavor branes is a good approximation only when $N_f/N_c \ll 1$; since we are in the large $N_c$ limit, flavor branes can be treated as probes as long as they are in finite number. On the gauge theory side, this amounts to consider finitely many flavors in 't Hooft's limit \cite{'tHooft:1973jz}: the leading order Feynman diagrams are planar diagrams without quarks running in internal loops. This is reminiscent of the `quenched' approximation in lattice gauge theories, where quantum dynamics is governed by the gluons, and quarks appear only as external fields.
In that limit, some interesting phenomena which are tight to the presence of quark bubbles in the QCD sea - screening of color charges and flux tube breaking for instance - are missed at the leading order. To account for such phenomena, Veneziano proposed a different scaling limit where the ratio $N_f/N_c$ is kept fixed \cite{Veneziano:1976wm}: the leading order Feynman diagrams are now planar diagrams where quarks are allowed to run in internal loops instead of gluons (with cubic vertices). On the dual gravity + branes side, the backreaction of flavor branes has now to be taken into account \cite{Casero:2006pt}.

By considering the addition of supersymmetric backreacting noncompact D7-branes to the KS solution, we will study a flavored version of the dual field theory in the leading order of Veneziano's expansion. The IR regime of the gauge theory is expected to describe a version of large $N$ SQCD with a quartic superpotential.


\section{KS field theory with nonchiral flavors}

By considering $N$ regular D3-branes at the tip of the singular conifold (described algebraically by the equation $z_1 z_2-z_3z_4=0$ in $\mathbb{C}^4$) and taking the decoupling limit, Klebanov and Witten \cite{Klebanov:1998hh} were able to find an $AdS^5\times T^{1,1}$ background with constant dilaton and $N$ units of 5-form flux. $T^{1,1}$ is a 5d Sasaki-Einstein manifold, the real cone over which is the singular conifold. The solution is dual to a 4d \Nugual{1} SCFT describing the low energy dynamics on the branes. It has gauge group $SU(N)\times SU(N)$ and bifundamental chiral superfields $A_i$ and $B_j$ ($i,j=1,2$), transforming in the $(\square,\overline\square)$ and $(\overline\square, \square)$ representations of the gauge groups, coupled through an $SU(2)\times SU(2)\times U(1)_B\times U(1)_R$ quartic superpotential:
\begin{equation}
W = \lambda \,Tr(A_i B_k A_j B_l)\, \epsilon^{ij} \epsilon^{kl}\;.
\end{equation}
Klebanov and Tseytlin \cite{Klebanov:2000nc} (KT) analyzed the addition of $M$ fractional branes to this setup: the result is that the rank of one of the two gauge groups becomes $SU(N+M)$, and the field theory is no longer conformal. They found a dual supergravity solution with constant dilaton, nontrivial 3-fluxes and running 5-flux, which displays a repulson-like singularity in the IR. Klebanov and Strassler \cite{Klebanov:2000hb} understood that the RG flow of such field theory undergoes an infinite cascade of Seiberg dualities, where ranks change according to $\ldots\to SU(N+M)\times SU(N)\to SU(N-M)\times SU(N)\to SU(N-M)\times SU(N-2M)\to\ldots$. If $N=kM$, at the end of the cascade, in the symmetric point of the baryonic branch of the moduli space of the $SU(2M)\times SU(M)$ theory, the low energy physics is essentially the one of \Nugual{1} pure SYM. Quantum deformation of the moduli space at the last step of the cascade implies that the relevant CY space is not the singular conifold, but rather the deformed conifold ($z_1 z_2-z_3z_4=\epsilon$ in $\mathbb{C}^4$), which is endowed with a finite volume $S^3$ at the tip. The complex structure deformation parameter is also related to the gluino condensate of the low energy \Nugual{1} pure SYM.
With this picture in mind, Klebanov and Strassler were able to find a smooth supergravity solution, asymptoting KT for large values of the holographic coordinates, with 3-form fluxes and running 5-form flux on the deformed conifold. That solution describes at low energy \Nugual{1} pure glue theory, and the cascade is a purely field theoretical UV completion which keeps the gauge theory always strongly coupled, and the dual gravity background weakly curved.

\begin{figure}
\begin{minipage}{72mm}
\includegraphics[width=\linewidth,height=30mm]{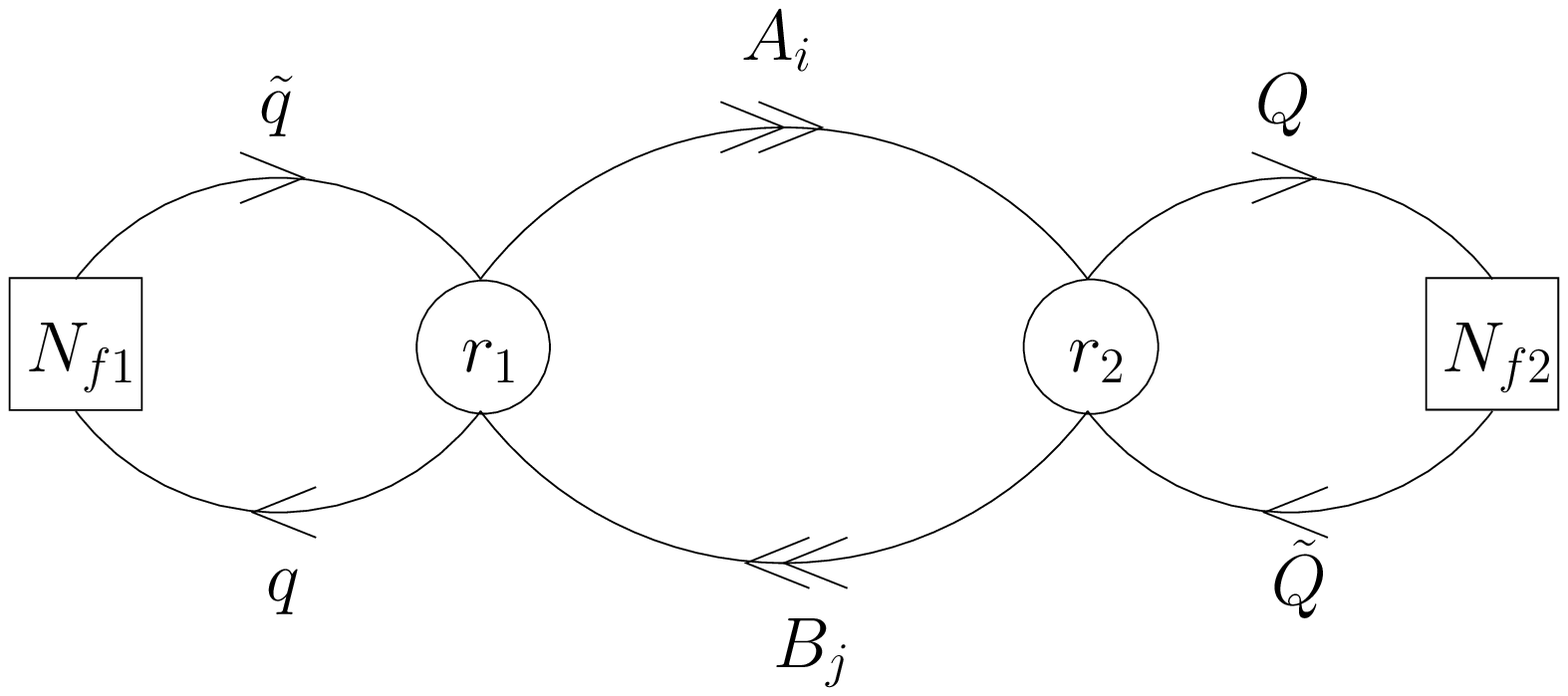}
\caption[quiverN]{{The quiver diagram of the gauge theory. Circles are gauge groups, squares are flavor groups, and arrows are bifundamental chiral superfields. $N_{f1}$ and $N_{f2}$ sum up to $N_f$. \label{quiverN}}}
\end{minipage}
\hfil
\begin{minipage}{65mm}
\includegraphics[width=\linewidth,height=35mm]{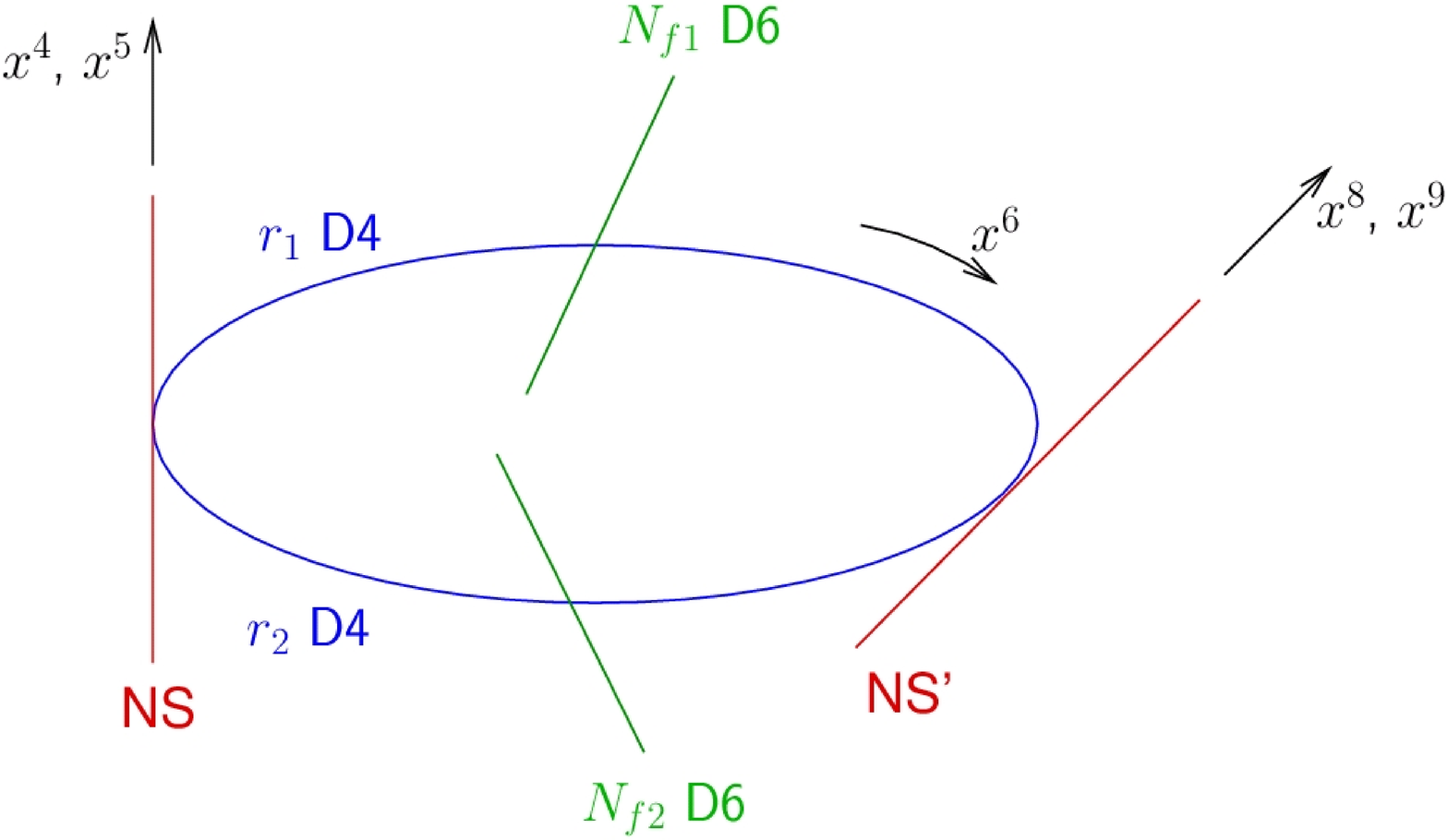}
\caption[IIA]{Type IIA Hanany-Witten construction of the flavored field theory.\label{IIA}}
\end{minipage}
\end{figure}

Addition of supersymmetric flavor branes to the KS background is clearly an important step towards finding a string dual of strongly coupled, large $N$ SQCD. In \cite{Benini:2007gx}, the present author and collaborators found an analytic solution accounting for unquenched massless flavors, realized in the dual background by means of backreacting noncompact D7-branes. We started from the holomorphic embedding $z_1+z_2=0$ \cite{Kuperstein:2004hy} for the D7. This embedding is $\kappa$-symmetric and its CY part contains a two-cycle, which is exceptional in the case of the singular conifold and of finite size in the case of the deformed conifold. The D7-brane can be equipped or not with one unit of worldvolume gauge flux on this 2-cycle, resulting into nonchiral flavors for one or the other gauge group respectively.
The field content of the flavored gauge theory is encoded in the quiver diagram of Fig. \ref{quiverN}.
By means of the T-dual type IIA Hanany-Witten setup drawn in Fig. \ref{IIA}, one can show that the flavors interact also through the tree level superpotential (traces are understood)
\begin{equation}
\begin{split} \label{superpotential}
W & = \lambda (A_1 B_1 A_2 B_2 -A_1 B_2 A_2 B_1) + h_1 \, \tilde{q} (A_1 B_1 + A_2 B_2)q + h_2 \, \tilde{Q} (B_1 A_1 + B_2 A_2)Q + \\
&\quad +\alpha \, \tilde{q} q \tilde{q} q + \beta \, \tilde{Q} Q \tilde{Q} Q  \;.
\end{split}
\end{equation}

\section{The dual supergravity + branes background}

To give a dual description of the aforementioned field theory in the Veneziano limit, in \cite{Benini:2007gx} we have looked for a supersymmetric solution of the equations of motion arising from the action $S=S_{IIB}+S_{D7}$, where $S_{IIB}$ is the action of type IIB supergravity and $S_{D7}$ is the Dirac-Born-Infeld plus Wess-Zumino action of the flavor branes. The solution is characterized by the metric, the dilaton, RR and NSNS 3-form field strengths $F_3$ and $H_3$, RR 1-form field strength $F_1$, and the D7-branes embedding. D7-branes are magnetic sources for $F_1$, violating its Bianchi identity, and source the dilaton and Einstein equations of motion.

In the previous discussion, we have considered the D7 embedding $z_1+z_2=0$. It would be nice to find a backreacted solution for flavored branes with that localized embedding. Unfortunately, because this specific embedding explicitly breaks the $SU(2)\times SU(2)$ isometry of the (deformed) conifold, that is a very hard task.
Therefore we take advantage of an available trick: acting with the broken $SU(2)\times SU(2)$ symmetry on the embedding $z_1+z_2=0$, we can find a continuous class of mutually BPS embeddings of the flavor branes, and \emph{smearing} the sources, namely summing homogeneously over this whole class of flavor brane sources, we are able to recover the continuous isometries of the unflavored background. The action of smearing on the gravity + branes side is that of transforming the 8-dimensional localized brane action into a 10-dimensional integral; for instance, the leading term in the WZ action becomes $S_{D7}^{WZ}\supset T_7 \int\Omega\wedge C_8$, where $\Omega=-\frac{N_f}{4\pi} \sum_i \sin\theta_i d\theta_i\wedge d\varphi_i$ is a manifestly $SU(2)\times SU(2)$-invariant 2-form. Similar effects occur for the DBI action. On the field theory side, the effect is that of `smearing' over the flavor-bifundamental quartic superpotential \eqref{superpotential}. A thorough analysis of the effects of the smearing, on both sides of these dualities, can be found in \cite{Benini:2006hh}. For the time being, it suffices to say that these effects do not enter in the most important physical phenomena exhibited by our solution, such as the duality cascade and the IR dynamics, where supposedly the bifundamental fields disappear, as in the unflavored solution.

By exploiting the symmetries of the problem, we can write an educated \emph{ansatz} which also solves the (modified) Bianchi identities; solution of DBI equations for the flavor branes embeddings follows from $\kappa$-symmetry; finally, the dilaton and Einstein equations are reduced to a BPS system of first order ODEs by requiring supersymmetry of the solutions.
The ansatz and an analytic solution of the BPS equations, that we do not reproduce here for lack of space, can be found in \cite{Benini:2007gx}. The solution looks at first sight quite similar to the one of KS for the metric and flux part, with two notable differences: a nontrivial $F_1$ and a running dilaton
\begin{equation}
e^{\phi(\tau)}=\frac{4\pi}{N_f} \frac{1}{\tau_0 -\tau}\;.
\end{equation}
$\tau$ is the usual radial coordinate of the deformed conifold, which is $0$ at the tip, and $\tau_0$ is a positive integration constant. Notice that then the range of the radial coordinate is $0\leq \tau< \tau_0$: the interpretation of this maximal value of $\tau$ will be discussed shortly.
Consistently, the unflavored KS solution \cite{Klebanov:2000nc} can be obtained in the limit $N_f\to 0$, with $N_f \tau_0$ fixed in order to find a finite (and constant) dilaton.

The background we have found has two curvature singularities: a UV singularity and an IR one. The first one is found when $\tau$ approaches its maximal value $\tau_0$ and is the usual singularity that occurs in systems with backreacting 7-branes, found also in \cite{Benini:2006hh}. The physical interpretation of this singularity in our solution is the appearance of a Landau pole for the `diagonal' gauge coupling $(g_1^{-2} + g_2^{-2})^{-1/2}$. On top of this, also a `duality wall' appears. We will discuss this point later.

The second singularity is an IR singularity: when $\tau\to 0^+$, all the components of the metric (as well as the 3- and 5-fluxes and the dilaton) look like in the unflavored KS case, but there are $\mathcal{O}(\tau)$ instead of $\mathcal{O}(\tau^2)$ corrections. The fibration of $S^3$ over $\mathbb{R}^3$ becomes singular. 
Were not for this curvature singularity, the geometry would have ended smoothly, and linear confinement (infinitely long QCD strings) would have been found by studying the behavior of spatial Wilson loops. In Veneziano's limit, pair production of flavor fields is allowed already at the leading order, leading to charge screening and flux tube breaking. Hence it is natural to expect a dual solution with a curvature singularity in the IR region. We have checked, by means of Maldacena's method for computing the Wilson loop, that exactly the $\mathcal{O}(\tau)$ corrections which are responsible for this curvature singularity are also responsible for the existence of a maximal distance up to which a pair of external quarks and antiquarks can be separated while keeping a single flux tube in between.


\section{RG flow of the flavored theory: duality cascade and duality wall}

Let us speculate on the possible RG flow of the flavored theory. Assume that at some energy scale the addition of flavors does not spoil the property of the KS solution of having the gauge group of larger rank running towards weak coupling, and the gauge group of smaller rank running to weak coupling. We will see shortly that this assumption is met by our solution at least in the UV, where ranks satisfy the hierarchy $r_{1,2}\gg |r_1-r_2|\gg N_{f1,f2}$. Then, flowing towards the IR, at some energy scale the gauge coupling of the largest group diverges, and we have to resort to a Seiberg dual description, as in the unflavored case. The result is that the rank of the larger gauge group diminishes, and becomes the smaller of the two. The flavor groups remain untouched, but the one which was coupled to the larger gauge group now is coupled to the smaller one. A key point is that the flavored field theory with the field content of Fig. \ref{quiverN} and the quartic superpotential \eqref{superpotential} is \emph{self-similar}. The gauge theory resulting out of a Seiberg duality has the \emph{same} field content and superpotential as before the duality: only the ranks of one gauge group shifts by an integer, and the nonzero couplings are reshuffled. Self-similarity is a property of the quartic superpotential: it does not hold for chiral flavors such as those considered in \cite{Ouyang:2003df}, which are coupled to bifundamental fields through a trilinear superpotential. In that case, as discussed in \cite{Benini:2007kg}, if one takes into account additional gauge singlet modes arising at the intersection of D7 with flux, the theory is almost self-similar: there is also a superpotential coupling between gauge singlet, flavors and bifundamentals, and only the power of bifundamental fields changes under Seiberg duality.

Going back to our flavored theory, as long as the assumptions are met, we expect a cascade of dualities to be at work. But we can go beyond these expectations: equipped with our supergravity solution, we are able to extract from it the RG flow of the dual field theory, at least in the UV regime, where the cascade is well under control on the supergravity side. In \cite{Benini:2007gx}, two solutions were actually found. The first one is that being discussed here: it enjoys a duality cascade from the UV until the IR, where nonperturbative gauge dynamics, supposedly quantum deformation of the moduli space at the last step of cascade, occurs. If that is the case, after selecting the most symmetric point of the baryonic branch of this moduli space, the very low energy dynamics of the dual field theory may be \Nugual{1} SQCD with a quartic superpotential, somehow similarly to \cite{Casero:2006pt}.
The second solution displays the same kind of duality cascade in the UV, but the initial conditions on the ranks are such that below some energy scale, after a last Seiberg duality, both gauge groups flow to weak coupling: this solution asymptotes the IR regime of the corresponding nonchirally flavored KW theory, whose supergravity background is after the smearing the same as \cite{Benini:2006hh}, where originally chiral flavors were considered.

In order to read cleanly the RG flow from our solution, we recall that, since the gauge theory can be microscopically engineered by means of fractional D3 and D7 branes, and given the radius-energy correspondence, it is natural to associate brane charges of the background at a given value of $\tau$ to ranks of groups at the dual energy $E$. This is standard reasoning in the gauge/gravity literature of duality cascades.  
Usually, in the literature, the brane charges which are computed in the background are fluxes of the gauge-invariant improved RR field strengths, which according to the classification of  \cite{MarolfCB}  are the so called `Maxwell charges': they are gauge invariant, conserved, carried by fluxes and in general not quantized. In our case the D7-charge (integral of $d F_1$) is constant and equal to $N_f$, the total number of flavors, whereas the D5 and D3-charges (fluxes of $F_3$ and $F_5$ respectively) run. 
At \emph{any} energy scale $E$, one can consider the system of branes which engineers the field theory under study: stacks of wrapped D5 branes with or without worldvolume flux and of D7 branes with or without worldvolume flux on the 2-cycle in their worldvolume. Computing the total (gauge invariant) `brane source charges' of this system, that can be read from the WZ action and depends on the integral of $B_2$ on the 2-sphere at the value of $\tau$ dual to $E$, and matching these total charges with the Maxwell charges, it is possible to extract the values of the ranks of gauge and flavor groups at \emph{any} energy scale. Consistently with the field theory picture, they remain constant in each Seiberg duality cascade step, and change discontinuously when a Seiberg duality has to be performed. 

This procedure for extracting ranks is correct, but quite involved. The reason is that we are using brane charges which are not quantized. It is much more natural to directly identify ranks with quantized brane charges. As understood in \cite{Benini:2007gx}, this possibility is offered by a third definition of brane charges that can be formulated in theories with Chern-Simons terms, namely the so called `Page charges' \cite{Page,MarolfCB}, which involve $B_2$ together with RR improved field strengths: by construction, these charges are conserved, carried by fluxes, and gauge invariant under small gauge transformations but not under large ones; hence they are quantized, as monopole numbers.
Repeating the matching described above, but now using Page charges on one hand and brane source charges (but putting $B_2=0$) on the other hand, one can very directly extract the unknown ranks from the integer Page charges that are measured in the gauge where $b\equiv \frac{1}{4\pi^2 \alpha'}\int_{S^2}B_2\in[0,1]$.

At the same time, thanks to the gauge/gravity asymptotic dictionary relating the gauge couplings with the dilaton and the $B_2$ integral (in the gauge where $b\equiv \frac{1}{4\pi^2 \alpha'}\int_{S^2}B_2\in[0,1]$)
\begin{equation}\label{dictionary}
\frac{4\pi^2}{g_l^2}+\frac{4\pi^2}{g_s^2} = \pi e^{-\phi} \quad, \qquad\qquad \frac{4\pi^2}{g_l^2}-\frac{4\pi^2}{g_s^2} = 2\pi e^{-\phi} \left(b -\frac{1}{2}\right)\;,
\end{equation}
we can follow the RG flow of the couplings, up to the unknown radius-energy relation. 
A Seiberg duality is needed every time $b(\tau)$ reaches an integer value ($0$ in the gauge used above). Correspondingly, a large gauge transformation bringing $b$ from 0 to 1 is needed in order to use the same dictionary \eqref{dictionary}. 

Summarizing the point, we are led to the very interesting and straightforward picture where ranks can be computed by means of Page charges in the gauge where $b\in[0,1]$, and Seiberg duality can be seen as a large gauge transformation on the dual supergravity side.

Both in the involved `Maxwell' frame and in the simpler and equivalent `Page' frame, everything works consistently; similarly, the R-anomalies can be successfully matched on the two sides of the duality.

Up to small and qualitatively irrelevant corrections stemming from departure of the radius/energy relation from the conformal one, the RG flow in the UV region is depicted in the bottom right part of Fig. \ref{RGflows}: blue lines represent the inverse squared gauge couplings, whereas the red line is their sum. At the energy scale dual to $\tau_0$ the sum of the inverse square gauge coupling vanishes: this is a Landau pole. On top of this, $b$ diverges at $\tau_0$, meaning that it is also an accumulation point of the energies where a Seiberg duality is needed. This phenomenon is called `duality wall' \cite{Strassler:1996ua}. To our knowledge, this is the first explicit realization of this phenomenon in a gravity dual.

Finally, we can compare how gauge couplings run in the KW solution \cite{Klebanov:1998hh}, KS solution \cite{Klebanov:2000hb}, flavored KW solution \cite{Benini:2007kg} and flavored KS solution \cite{Benini:2007gx}, see again Fig. \ref{RGflows}. 
\begin{vchfigure}[h]
\includegraphics[width=.75\textwidth]{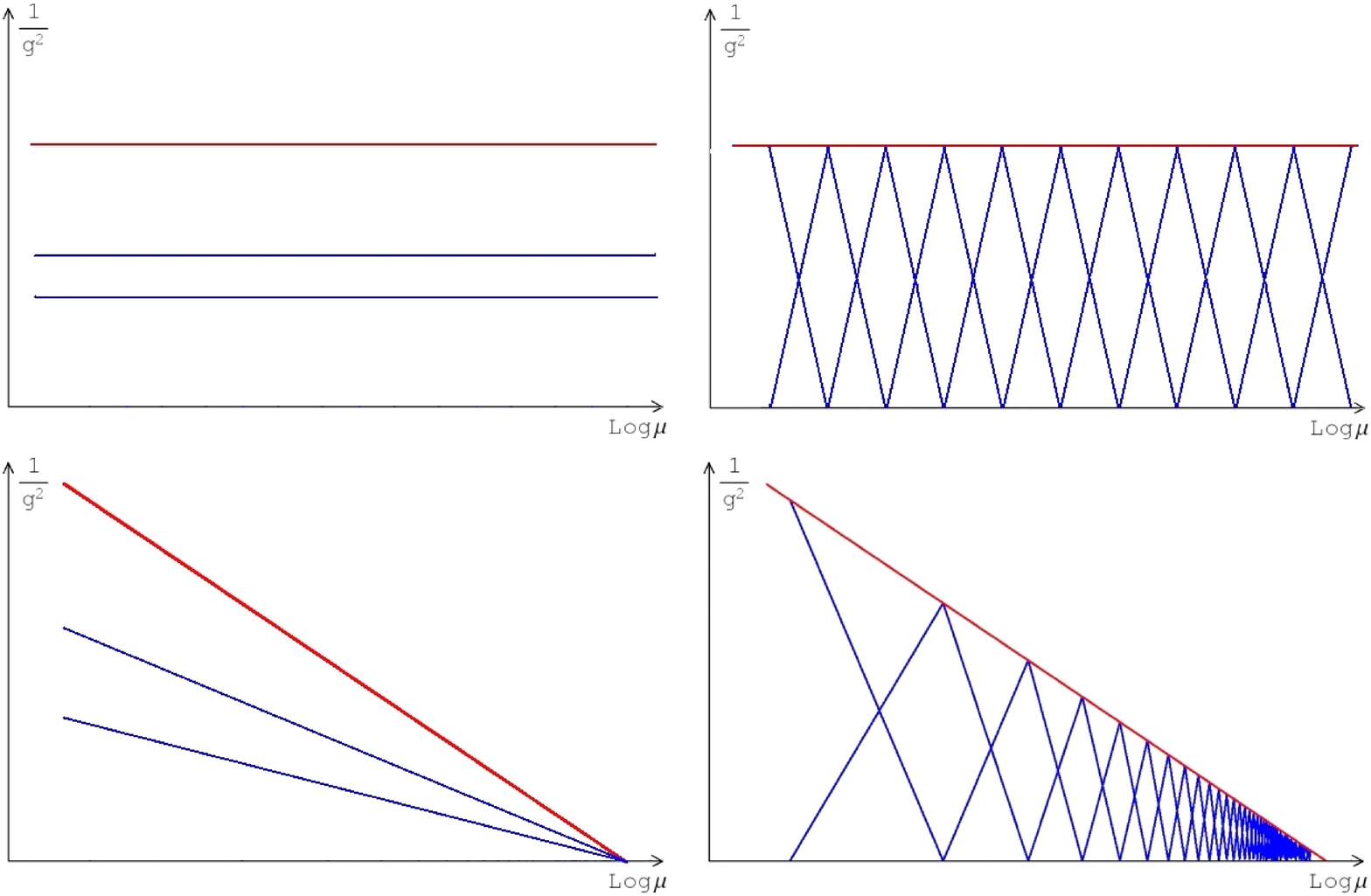}
\vchcaption{RG flows of KW (top left), KS (top right), flavored KW (bottom left), flavored KS (bottom right).}
\label{RGflows}
\end{vchfigure}
Adding fractional D3 branes to the conformal KW solution causes the difference of the inverse squared gauge coupling to run and creates an infinite Seiberg duality cascade ($b\to\infty$ when $\tau$ approaches its maximal value $\infty$). Adding flavor D7 branes causes the sum of the inverse squared gauge coupling to run, and creates a Landau pole at a finite UV energy scale. Pictorially, the flavored KW RG flow can be obtained from the trivial KW one by pinching the point at infinity and pushing it to a finite value.  The flavored KS solution reduces to the flavored KW solution in the limit where fractional D3 branes disappear, and to the unflavored KS solution in the scaling limit $N_f\to 0$, $N_f \tau_0$ fixed. An analogous limit reduces the flavored KW to the KW solution. Going in the opposite direction, superposing the two effects leads to an expected RG flow for the flavored KS theory which looks like the observed one.

%
%
%
%


\def\bstname{fdp}

\end{document}